\documentclass{ifacconf}

\usepackage{graphicx}      % include this line if your document contains figures
\usepackage{natbib}

\usepackage{enumerate}
\usepackage{amsmath}
\usepackage{amscd}
\usepackage{amssymb}
\usepackage{latexsym}
\usepackage{array}
\usepackage{tabularx}
\usepackage[ruled]{algorithm}
\usepackage{algorithmic}
\usepackage{epsfig,subfigure}

\usepackage{multirow}
\usepackage{bigstrut}
\usepackage{array}
\usepackage{epstopdf}
\usepackage{graphicx}

\newcommand{\PreserveBackslash}[1]{\let\temp=\\#1\let\\=\temp}
\newcolumntype{C}[1]{>{\PreserveBackslash\centering}p{#1}}
\newcolumntype{R}[1]{>{\PreserveBackslash\raggedleft}p{#1}}
\newcolumntype{L}[1]{>{\PreserveBackslash\raggedright}p{#1}}

\newtheorem{deft}{Definition}

\begin{document}

\begin{frontmatter}

\title{A Robust Compressive Quantum State Tomography Algorithm Using ADMM\thanksref{footnoteinfo}}

\thanks[footnoteinfo]{This work was partially supported by the Swedish Research Council, the Linnaeus Center ACCESS at KTH, the European Research Council under the advanced grant LEARN, contract 267381, and the China National Key Basic Research Program under Grant No. 2011CBA00200}

%\author[First]{1}
%\author[Second]{2}
\author[First]{Kezhi Li}
\author[Second]{Shuang Cong}
%\author[Third]{Third C. Author}

\address[First]{ACCESS Linnaeus Centre, Royal Institute of Technology (KTH),\\
SE 10044 Stockholm, Sweden (e-mail: kezhi@kth.se).}
\address[Second]{Dept. of Automation, Univ. of Science and Technology of China,\\
   Hefei, 230027, China (Corresponding author’s email: scong@ustc.edu.cn)}
%\address[Third]{Electrical Engineering Department,
%   Seoul National University, Seoul, Korea, (e-mail: author@snu.ac.kr)}

\begin{abstract}                % Abstract of not more than 250 words.
    The possible state space dimension increases exponentially with respect to the number of qubits. This feature makes the quantum state tomography expensive and impractical for identifying the state of merely several qubits. The recent developed approach, compressed sensing, gives us an alternative to estimate the quantum state with fewer measurements. It is proved that the estimation then can be converted to a convex optimization problem with quantum mechanics constraints. In this paper we present an alternating augmented Lagrangian method for quantum convex optimization problem aiming for recovering pure or near pure quantum states corrupted by sparse noise given observables and the expectation values of the measurements. The proposed algorithm is much faster, robust to outlier noises (even very large for some entries) and can solve the reconstruction problem distributively. The simulations verify the superiority of the proposed algorithm and compare it to the conventional least square and compressive quantum tomography using Dantzig method.
\end{abstract}


\end{frontmatter}

\section{Introduction}

The interests of applying control theory and signal processing techniques to quantum mechanics have increased dramatically in recent decade. One objective is to develop a series of systematic methods for the active manipulation and control of quantum systems. The foundation of such theory lies in the fact that we are able to prepare and measure a given quantum state efficiently. It is not trivial since the microscopic quantum systems have their unique features, on account of which they significantly differ from the classic world. In practice, people often use the measurement data to estimate an unknown quantum state. In mathematics, a quantum pure state $|\psi \rangle$ can be described as a vertical vector with size $d$ in Hilbert space. This vector is called a state vector and it theoretically contains statistical information about the quantum system. For the mixed state that corresponds to a probabilistic mixture of pure states, a state vector is not enough. It usually requires a $d \times d$ density matrix $\mathbf{\rho}$ to depict the quantum state by giving the probabilities in each possible state, which implies that $O(d^2)$ parameters are needed to describe an arbitrary state in a $d$ dimensional Hilbert space.

However in fact, most states people are interested in real life are nearly pure. Here nearly pure means the quantum state is a mixed state that can be represented as the probabilistic combination of equal to or less than $r$ pure states. Suppose the unknown mixed state is a probabilistic mixture of $r \sim O(1)$ pure states, then it means its density matrix $\mathbf{\rho}$ has rank not larger than $r$. This pre-information enables us to reduce the number of parameters to identify a quantum system (\cite{Gross-QuantumState}). By using a novel signal processing technique called compressed sensing (CS) (\citet{donoho-cs})
that has been widely investigated in last a few years, people are able to obtain good estimates of nearly pure quantum states with $O(r d \log d)$ expectations and corresponding observables (\cite{Gross-QuantumState,Liu-Universal-Pauli}). Thus the required number of identifying a quantum state can be reduced dramatically by solving an optimization problem when $d$ goes to large, and its effectiveness has been verified by a series of experiments such as in \cite{Smith-QuantumState}. While in current literatures of quantum state tomography via CS, this problem cannot be efficiently handled by generic optimization solvers because of the large number of involving variables. For instance in \cite{Smith-QuantumState} the authors summarized the estimation to least squares (LS) problem or compressed sensing (CS) problem, and solved them by using the prevalent convex optimization toolbox. This paper addresses the problem of state recovery using low rank information. Our work is inspired by the an alternating augmented Lagrangian method (ADMM) which has received much attentions from the optimization community (\cite{Boyd-ConvexOpt,Boyd-DistributedOpt}) though it was developed in the 1970s. In this paper we reformulate the quantum state tomography to a optimization problem and design a fast algorithm based on ADMM where in each step trying to optimize the density matrix and project it onto the constraint set of quantum states according to the Karush–-Kuhn--Tucker (KKT) conditions, and the algorithm finally reaches a solution with good accuracy. Due to the alternating properties of the algorithm, we are able to run it distributively, and normally obtain  the quantum state with most purified result by minimizing the nuclear norm of the density matrix, a heuristic for minimizing the rank (\cite{Candes-Tight}). People have proved many matrices bases including Pauli matrices satisfy the rank restricted isometry property (RIP) introduced in \cite{RECHT-Guaranteed} with regard to low rank recovery such that they are capable to recover the unique density matrix with sufficient measurements via compressive sensing approach. This work can be deemed as the reconstruction part of the compressive quantum state tomography, which gives a solution to identifying the density matrices accurately and efficiently for standard tomography as well as continuous tomography (\cite{Smith-QuantumState}).

This paper is organized as follows. In Section II, we will explain the idea of quantum state tomography via compressive sensing and the framework of ADMM. In Section III the compressive state tomography with quantum constraints is formulated formally, and the proposed algorithm is introduced and analyzed in detail. Simulations verify the effectiveness of the proposed approach in Section IV, and finally the conclusion is summarized in Section V. Moreover, some necessary supplementary knowledge is explained in the Appendix.

\textbf{Notation}: Bold letters are used to denote a vector or a matrix. For vectors, $||\cdot||_1, ||\cdot||_2$ represent the $l_1, l_2$ norm, respectively. For matrix, $\mathbf{A}^T$ and $\mathbf{A}^*$ denote the transpose and Hermitian transpose of $\mathbf{A}$, respectively. $||\cdot||_p$ denotes the Schatten $p$-norm with $||\mathbf{A}||_p= \left( \sum_i \sigma_i(\mathbf{A}^p)^{1/p}\right)$, where $\sigma_i(\mathbf{A})$ are the singular values of $\mathbf{A}$. Specifically, $||\cdot||_*$ is the nuclear norm and $||\cdot||_F$ represents the Frobenius norm. $\text{tr}(\cdot)$ is the operator to calculate the trace. $\text{vec}(\mathbf{A})$ represents the vertical vector concatenates $\mathbf{A}$'s columns, and ``$\text{mat}$" is its inverse operator to convert a vector to a matrix. Bra-–ket notations $|\psi_i \rangle$ are used to denote quantum states. $\mathbf{A} \succeq 0$ means $\mathbf{A}$ is a positive semi-definite matrix.

\section{Compressive Quantum State Tomography and ADMM}

\subsection{Quantum State Tomography Via Compressive Sensing}

The task of quantum state tomography is to reconstruct the quantum states processed and produced by physical systems. Due to the special characteristics of the quantum mechanics, a $d \times d, d=2^q$ density matrix $\mathbf{\rho}$, a quantum-mechanical analogue to a phase-space probability measure, is used to describe a quantum system. Since the degree of the freedom of $\mathbf{\rho}$ is $d \times d$, usually people need the number of measurements increase with exponential growth regarding the state space dimension $d$ in order to identify $\mathbf{\rho}$. If we make the measurements discretely and denote the observable matrix $\mathbf{O}_i$, the the expectation of measurements $\mathbf{y}_i \in \mathbb{R}^m$, and measuring operator $\mathcal{A}: \mathcal{C}^{d\times d \rightarrow m}$, then
\begin{equation}\label{eq:y=Arho+e}
\begin{split}
\mathbf{y}_i= (\mathcal{A}(\rho))_i + e_i&= c \text{tr}(\mathbf{O}_i^* \rho) +e_i, \ \ \ i=1, \cdots, m, \ \ \ \text{or} \\
\mathbf{y} &= \mathbf{A} \text{vec} (\mathbf{\rho}) +\mathbf{e},
\end{split}
\end{equation}
where $\mathbf{A}\in \mathbb{C}^{m \times d^2}$ is the normalized operator whose $i$th row is the concatenation of $\mathbf{O}_i^*$'s rows, $\mathbf{e}  \in \mathbb{R}^m$ represents the noise due to the system or measuring process. $c$ is some normalized constant. If we set $E(\mathcal{A}^* {\mathcal{A}}) = \mathcal{I}$, $c$ would be $\frac{d}{\sqrt{m}}$. Conventionally, people use the least square approach to estimate $\rho$
\begin{equation}\label{eq:rho=argmin}
\begin{split}
\hat{\rho} = \arg \min_{\rho} \sum_i{\left[ \mathbf{y}_i - c \text{tr}(\mathbf{O}^*_i \rho)\right]^2 },     \\
\text{s.t.} \ \ \rho^* = \rho, \ \rho \succeq 0, \  \text{tr}(\rho)=1.
\end{split}
\end{equation}
Because the degrees of freedom of $\rho$ is $O(d^2)$, normally $O(d^2)$ measurements are needed to identify the unique state.

Yet assuming the underlying quantum system is pure or nearly pure, $\mathbf{\rho}$ becomes a probabilistic weighted combination of equal to or less than $r$ rank-$1$ matrices derived from a series pure states (see details in Appendix). When $r$ is small, people have suggested that $O(rd \log d)$ settings would possibly suffice instead of $d^2$. Minimizing the rank of a matrix belongs to NP-hard problems, so alternatively people pursuit the solution by minimize the nuclear norm of density matrix $||\rho||_*= \operatorname{tr} \left(\sqrt{\rho^*\rho}\right) = \sum_{i=1}^{\min\{m,\,n\}} \sigma_i$, which is a convex function that can be optimized efficiently. The nuclear norm has been proved as the best convex approximation of the rank function over the unit ball (\cite{RECHT-Guaranteed}), so minimizing $||\rho||_*$ is a heuristic for minimizing the rank (\cite{Gross-QuantumState}). Thus $\rho$ with low rank can be estimated by compressed sensing approaches such as in \cite{Liu-Universal-Pauli}
\begin{equation}\label{eq:Danzig_LASSO}
\begin{split}
&\text{Dantzig with quantum constraints:} \ \\
& \hat{\rho} = \arg \min_{\rho} ||\rho||_* \ \\
 &\text{s.t.} \ \sum_i{\left[ \mathbf{y}_i - c \text{tr}(\mathbf{O}^*_i \rho) \right]^2 \leq \epsilon, \rho^* = \rho, \ \rho \succeq 0}, \ \ \ \text{or}\\
&\text{LASSO with quantum constraints:} \ \\
& \hat{\rho} = \arg \min_{\rho}{\frac{1}{2} ||\mathbf{y}_i - c \text{tr}(\mathbf{O}^*_i \rho)||_2^2 + \mu ||\rho||_*} \ \\
 &\text{s.t.} \ \rho^* = \rho, \ \rho \succeq 0,
\end{split}
\end{equation}
where $\epsilon, \mu$ are parameters. In this paper we develop a convex optimization algorithm based on ADMM to solve above problems corrupted by sparse outliers with quantum constraints.

\subsection{Alternating Direction Method of Multipliers (ADMM)}
ADMM is an optimization method with good robustness and can support decomposition. Consider the optimization problem such as
\begin{equation}\label{eq:min(f+g)}
\begin{split}
\text{minimize} \ \ f(\mathbf{x}) + g(\mathbf{z}) \ \ \text{s.t.} \ \ \mathbf{Ax}+ \mathbf{Bz} = \mathbf{c}
\end{split}
\end{equation}
for some variable $\mathbf{x},\mathbf{z} \in \mathbb{R}^n$, where $f, g: \mathbb{R}^n \rightarrow \mathbb{R}$ are convex functions. The augmented Lagrangian of (\ref{eq:min(f+g)}) is defined
\begin{equation}
L_\lambda(\mathbf{x},\mathbf{z},\mathbf{y}) = f(\mathbf{x}) + g(\mathbf{z}) + \mathbf{y}^T(\mathbf{Ax}+\mathbf{Bz}- \mathbf{c})+ \frac{\lambda}{2}||\mathbf{Ax}+\mathbf{Bz}- \mathbf{c}||_2^2.
\end{equation}
where $\lambda>0$ is a tunable parameter. Then the $k$th iteration of ADMM algorithm consists of three steps as follows
\begin{equation}\label{eq:ADMM_XZU}
\begin{split}
&1)\ \mathbf{x}^{k+1} = \arg\min_\mathbf{x} L_{\lambda}(\mathbf{x},\mathbf{z}^k,\mathbf{y}^k) \ \ \ \ \ \  // \ \mathbf{x}\text{-minimization}\\
&2)\ \mathbf{z}^{k+1} = \arg\min_\mathbf{z} L_{\lambda}(\mathbf{x}^{k+1},\mathbf{z}^k,\mathbf{y}^k) \ \ // \ \mathbf{z}\text{-minimization} \\
&3)\ \mathbf{y}^{k+1} = \mathbf{y}^{k} + \lambda(\mathbf{A}\mathbf{x}^{k+1} +\mathbf{B} \mathbf{z}^{k+1} - \mathbf{c})  \ \  \  // \ \text{dual-update}
\end{split}
\end{equation}

From above steps one can see that if we minimize over $\mathbf{x}$ and $\mathbf{z}$ jointly, the approach reduces to the classic method of multipliers. Instead, people split the augmented Lagrangian and minimize over $\mathbf{x}$ with $\mathbf{z}$ fixed and vice versa. The three steps are repeated until convergence. Certain stopping criteria is made to decide when the algorithm achieves a convergence. For instance, the algorithm is iterated until the primal and dual residuals are bounded
\begin{equation}\label{eq:stop_cri}
\begin{split}
||\mathbf{A}\mathbf{x}^k+\mathbf{B}\mathbf{z}^k-\mathbf{c}||_2 &\leq \varepsilon_{\text{pri}}, \ \\
||\mathbf{x}^k-\mathbf{x}^{k-1}||_2+||\mathbf{z}^k-\mathbf{z}^{k-1}||_2 &\leq \varepsilon_{\text{dual}},
\end{split}
\end{equation}
where $\varepsilon_{\text{pri}}>0, \varepsilon_{\text{dual}} >0$ are tolerance parameters. For more details and a complete convergence analysis, people who have interests may refer to \cite{Boyd-DistributedOpt}.

\section{Problem Formulation And Method}

In this section, we formulate the the problem of robust quantum state tomography and derive an efficient optimization algorithm using ADMM. Here ``Robust" means the algorithm fits for the circumstance the existence of not only small random noises, but also sparse outlier noises involved in the density matrix.

\subsection{Robust Compressive Quantum State Tomography}
During the measuring process of quantum state tomography, noises are involved due to the system or measurement errors.
%These errors can be divided to two categories. The first one is the normal noises $\mathbf{e}$ in (\ref{eq:y=Arho+e}).
Normally we assume $\mathbf{e}$ satisfying certain distribution (like Gaussian) and it can be minimized with least square techniques (\ref{eq:rho=argmin}), similar in Danzig or LASSO (\ref{eq:Danzig_LASSO}). However there exist abnormal circumstances in the measuring process that may cause the perturbation in the density matrix, and it can be reflected by sparse outlier entries in $\rho$ and of course not satisfying Gaussian distribution. We formulate these outlier entries as a sparse matrix $\mathbf{S} \in \mathbb{C}^{d\times d}$, then (\ref{eq:y=Arho+e}) becomes
\begin{equation}
\begin{split}
\mathbf{y}_i= (\mathcal{A}(\rho+ \mathbf{S}))_i + e_i&= c \cdot \text{tr}(\mathbf{O}_i^* (\rho+ \mathbf{S})) +e_i, \ \\
 \ i&=1, \cdots, m.
\end{split}
\end{equation}
In this case the result of least square method in (\ref{eq:rho=argmin}) will change significantly sometimes due to the existence of outliers. In addition, given the information that $\rho$ is relatively pure which implies it has low rank, the Dantzig/LASSO solver in (\ref{eq:Danzig_LASSO}) with low rank constraints based on truncated Singular Value Decomposition (SVD) might also fail because the sparse outliers effect the classic principle component analysis (PCA) dramatically in the process of dimensionality reduction. To reduce the influence of the noise to the rank estimation, we may reformulate the robust Dantzig solver with sparse outliers and quantum constraints to
\begin{equation}\label{eq:rho+S}
\begin{split}
&\text{minimize}\ \ ||\rho||_*+ ||\mathbf{S}||_1  \ \\
 &\text{s.t.} \  ||\mathbf{y} -  \mathbf{A} \text{vec}(\rho+\mathbf{S})  ||_2^2 \leq \epsilon, \rho^* = \rho, \ \rho \succeq 0,
\end{split}
\end{equation}
where $\mathbf{A}$ is with the same definition in (\ref{eq:y=Arho+e}). The idea of minimizing sparse noises can also be found in \cite{Candes-RobustPCA,Zhou-StablePri,Kyrillidis-MatriALSP} and has many applications in face recognition, etc. While in most previous papers the authors aimed for solving a matrix completion problem however here we want to recover the density matrix from observable measurements with special constraints on $\rho$. To involve the quantum constraints in ADMM, we re-write (\ref{eq:rho+S}) as
\begin{equation}\label{eq:rho+I+S}
\begin{split}
&\text{minimize}\ \ ||\rho||_*+ I_{\mathcal{C}}(\rho) +  ||\mathbf{S}||_1  \ \\
 &\text{s.t.} \  ||\mathbf{y} -  \mathbf{A} \text{vec}(\rho+\mathbf{S}) ||_2^2 \leq \epsilon,
\end{split}
\end{equation}
where $I_\mathcal{C}(\rho)$ is the indictor function on a convex set $\mathcal{C}$ with $I_\mathcal{C}(\rho)=0$ for $\rho \in \mathcal{C}$, and $I_{\mathcal{C}}(\rho) = \infty$ for $\rho \notin \mathcal{C}$, $\mathcal{C}(\rho)$ here is the Hermitian p.s.d. set satisfying $\rho^* = \rho, \ \rho \succeq 0$. So we have obtained two sets of variables with separable objective. The augmented Lagrangian can be derived as
\begin{equation}\label{eq:L_lambda_1}
\begin{split}
L_{\lambda_1}(\mathbf{\rho},\mathbf{S},\mathbf{u}') &= \left( ||\rho||_*+ I_{\mathcal{C}}(\rho) \right) +  ||\mathbf{S}||_1  \\
&+\mathbf{u}'^T(\mathbf{A}\text{vec}(\mathbf{\rho}) + \mathbf{A} \text{vec}(\mathbf{S})- \mathbf{y})\\
&+ \frac{\lambda_1}{2}||\mathbf{A}\text{vec}(\mathbf{\rho}) + \mathbf{A} \text{vec}(\mathbf{S})- \mathbf{y}||_2^2,
\end{split}
\end{equation}
where $\lambda_1$ is a parameter that can effect the rate of convergence and the number of iterations required. Or we may combine the linear and quadratic terms in (\ref{eq:L_lambda_1}) and it becomes
\begin{equation}\label{eq:L_lambda_11}
\begin{split}
L_{\lambda_1}(\mathbf{\rho},\mathbf{S},\mathbf{u}) &= \left( ||\rho||_*+ I_{\mathcal{C}}(\rho) \right) +  ||\mathbf{S}||_1 \\
&+ \frac{\lambda_1}{2}||\mathbf{A}\text{vec}(\mathbf{\rho}) + \mathbf{A} \text{vec}(\mathbf{S})- \mathbf{y}+ \mathbf{u}||_2^2,
\end{split}
\end{equation}
with $\mathbf{u}=(1/\lambda_1)\mathbf{u}'$.

\subsection{ADMM Steps}
We carry out the following steps in each iteration to solve (\ref{eq:rho+I+S}).
\subsubsection{Step 1}
In the $\rho$ minimization step, we update low rank $\rho$ with fixed $\mathbf{S}, \mathbf{u}$.
\begin{equation}
\begin{split}
\rho^{k+1} := &{\arg \min}_{\rho} \left\{||\rho||_*+ I_{\mathcal{C}}(\rho)  \right. \\
 &\left. +\frac{\lambda_1}{2}||\mathbf{A}\text{vec}(\mathbf{\rho}) +\mathbf{A} \text{vec}(\mathbf{S}^k)- \mathbf{y}+ \mathbf{u}^k||_2^2  \right\}.
\end{split}
\end{equation}
First, we minimize the unconstrained quadratic function in terms of $\rho$. The analytic solution to least square estimation can be written as
\begin{equation}
\rho_1^{k+1} = \text{mat} \left( \left(\mathbf{A}^*\mathbf{A} \right)^{-1}\mathbf{A}^*\left( \mathbf{y} - \mathbf{u}^k - \mathbf{A}\text{vec}(\mathbf{S}) \right)  \right).
\end{equation}
Second, project $\rho_1^{k+1}$ to $\rho_2^{k+1}$ on to the constraints set $\mathcal{C}$ at the same time with low rank, \emph{i.e.}
\begin{equation}
\rho_2^{k+1} = \Pi_{\mathcal{C}}(\rho_1^{k+1}),
\end{equation}
where $\Pi_{\mathcal{C}}$ denotes the Euclidean projection onto $\mathcal{C}$ and at the same time with low rank. For the particular constraint set of quantum state, $\mathcal{C}$ is a proper cone of the Hermitian p.s.d. matrices. We will show the projection process in Section \ref{section:projection} with efficient approach.

\subsubsection{Step 2}
In the $\mathbf{S}$ minimization step, we update sparse matrix $\mathbf{S}$ with fixed $\mathbf{\rho}^{k+1}=\mathbf{\rho}^{k+1}_2, \mathbf{u}$.
\begin{equation}
\begin{split}
\mathbf{S}^{k+1} := &{\arg \min}_{\mathbf{S}} \left\{||\mathbf{S}||_1  \right. \\
 &\left. +\frac{\lambda_1}{2}||\mathbf{A}\text{vec}(\mathbf{\rho}^{k+1}) +\mathbf{A} \text{vec}(\mathbf{S})- \mathbf{y}+ \mathbf{u}^k||_2^2  \right\}.
\end{split}
\end{equation}
It is a conventional LASSO problem and can be solved by iterations. However here we avoid solving it by a sequence of convex programs and adopt the shrink operator defined previously to calculate a solution efficiently. In detail, the least square estimate $\mathbf{S}$ can be approximated by
 \begin{equation}
\mathbf{S}_1^{k+1} =  \text{mat}\left( \left(\mathbf{A}^*\mathbf{A} \right)^{-1}\mathbf{A}^*\left( \mathbf{y} - \mathbf{u}^k - \mathbf{A}\text{vec}(\mathbf{\rho}^{k+1}) \right)  \right),
\end{equation}
and then shrink the magnitude to achieve a sparse solution
\begin{equation}
\mathbf{S}_2^{k+1}=\mathcal{S}_{\tau'}(\mathbf{s}) = \text{sgn}[\mathbf{s}]\max(|\mathbf{s}|-\tau' \mathbf{1},\mathbf{0})
 \end{equation}
 where $\mathcal{S}$ is the shrink operator also explained in Section 3.3, $\mathbf{s}=\text{vec}(\mathbf{S}_1^{k+1})$, $\tau'$ is a shrink parameter depends on the sparsity level of $\mathbf{S}$.

 \subsubsection{Step 3}
 At last we proceed the dual update step
 \begin{equation}
 \mathbf{u}^{k+1} =  \mathbf{u}^{k} +(\mathbf{y}-\mathbf{A}\text{vec}(\mathbf{\rho}^{k+1}) -\mathbf{A} \text{vec}(\mathbf{S}^{k+1})).
 \end{equation}
 This step is to record the alternative update direction and contribute to the next step.

 \subsubsection{Stop Criteria and Parameter Settings}
 The algorithm follows the steps 1-3 to carry out the updating information iteratively. In practice, relatively small numbers
of iterations, like 30-40, are sufficient to achieve a good accuracy. There are several stopping criterions, e.g. adopting bounds in (\ref{eq:stop_cri}) we have
\begin{equation}
\begin{split}
&||\mathbf{y} -  \mathbf{A} \text{vec}(\rho^k+\mathbf{S}^k)  ||_2^2 \leq \varepsilon_1||\mathbf{y}||_2,  \\
&||\mathbf{\rho}^k-\mathbf{\rho}^{k-1}||_2\leq \varepsilon_2, \ \ \ ||\mathbf{S}^k-\mathbf{S}^{k-1}||_2\leq \varepsilon_3.
\end{split}
\end{equation}
where $\varepsilon_1,\varepsilon_2,\varepsilon_3$ are parameters need to be tuned. Some methods of tuning parameters of alternating direction methods are indicated in \cite{Yuan-SparseAnd,Candes-RobustPCA}.

 \subsection{Projection onto Constraint Set with Low Rank}\label{section:projection}

 We utilize a \emph{positive eigenvalue thresholding operator} $\mathcal{D}_{\tau}$ to calculate $\rho_2^{k+1}$. Let $\mathcal{S}_{\tau}: \mathcal{R}^d \rightarrow \mathcal{R}^d$ denote the shrink operator such that
\begin{equation}
\mathcal{S}_{\tau}(\mathbf{x}) = \text{sgn}[\mathbf{x}]\max(|\mathbf{x}|-\tau \mathbf{1},\mathbf{0})
 \end{equation}
 here $\mathbf{1}$ is a vector with all elements $1$. The definition also can be extended to the matrix form. Then the positive eigenvalue thresholding operator $\mathcal{D}_{\tau}$ is defined as
 \begin{equation}\label{eq:rho2}
\mathbf{\rho}_2^{k+1}= \mathcal{D}_{\tau}(\mathbf{\rho}_1^{k+1}) = \mathbf{V}\mathcal{S}_{\tau}(\mathbf{\Sigma}^+) \mathbf{V}^*
  \end{equation}
   where $\mathbf{\Sigma},\mathbf{V}$ are obtained from the eigenvalue decomposition of a symmetrized matrix $1/2(\rho_1^{k+1}+{\rho_1^{k+1}}^*)$,
\begin{equation}\label{eq:VSigmaV}
\mathbf{V}\mathbf{\Sigma}\mathbf{V}^* = 1/2(\rho_1^{k+1}+{\rho_1^{k+1}}^*),
\end{equation}
   $\mathbf{\Sigma}^+$ only keep the positive part of the eigenvalues where $\mathbf{\Sigma}^+ = \max(\mathbf{\Sigma}, \mathbf{0})$, $\mathcal{S}_{\tau}(\mathbf{\Sigma}^+)$ is a shrink operator on diagonal matrix $\mathbf{\Sigma}^+$ which has eigenvalues as entries, $\tau=1/\lambda_1$. This approach can be derived from its Karush-–Kuhn-–Tucker (KKT) conditions of the optimal projection from $\mathbf{\rho}_2^{k+1}$ to set $\mathcal{C}$ with least square errors. Taking the indicator function $I_{\mathcal{C}}(\mathbf{\rho})$ for instance, under mild assumptions on a proper cone $\mathcal{C}$ the KKT conditions of
\begin{equation}
\begin{split}
&\text{minimize} \ \ ||\bar{\mathbf{\rho}}- \mathbf{\rho}||^2_2 \\
&\text{s.t.} \ \  \bar{\mathbf{\rho}} \in I_\mathcal{C}
\end{split}
\end{equation}
 are given by
\begin{equation}
\begin{split}
&\bar{\mathbf{\rho}} \in    I_\mathcal{C}, \ \ \ \  \bar{\mathbf{\rho}}-\mathbf{\rho}=\mathbf{\theta}, \\
&\theta \in I_\mathcal{C}, \ \ \ \ \mathbf{\theta}^* \bar{\mathbf{\rho}} =0.
\end{split}
\end{equation}
The third term is because positive semideﬁnite cone is self-dual. Then the Euclidean projection can be derived by decomposing $\rho$ into the difference
of two orthogonal elements:  one with nonnegative eigenvalues and one with negative part. After that the shrink operator leads to a solution satisfying low rank constraints. In addition, if given the information that the objective quantum state is the probabilistic linear combination of less than or equal to $r$ pure states, we may project $\rho$ to the set of $r$-rank matrices by selecting the maximum $r$ positive eigenvalues in $\mathbf{\Sigma}^+$ in (\ref{eq:rho2}). For the details of the derivation the readers may refer to (\cite{Boyd-ConvexOpt}).

\textbf{{Remark}}:\\
1) Regarding the convergence of ADMM and error bounds of recovering low rank matrix from its measurements the readers may refer to \cite{Boyd-ConvexOpt,Boyd-DistributedOpt, Wright-CompressivePri,Lin-LinearizedAlt}. If there is no analytical solution to (\ref{eq:ADMM_XZU}), we may also use the semidefinite programs. Details and softwere can be found in \cite{Sturm-UsingSeD}.

2) In practice, the observable $\mathbf{O}_i$ is not necessary the tensor product of Pauli matrices. For instance, in \cite{Smith-QuantumState} the author developed a device to proceed the quantum state tomography by continuous measurements where $\mathbf{O}_i$ is affected by outer radio frequency magnetic fields. In this case we can still use the proposed algorithm to recover the quantum state, as long as that $\mathbf{O}_i$ satisfy the rank RIP and number of measurements are sufficient large. Regarding the details of rank RIP and number of measurements of compressive quantum tomography, please refer to the Appendix.

3) If the dataset is large, our algorithm equipped with ADMM technique can be extended to a distributed manner as a consensus optimization problem. Assume the $N$ agents can communicate with each other, denote each cost function $f_i(\cdot)$, $i=1,2,\cdots$ as in (\ref{eq:y=Arho+e}), in this case (\ref{eq:ADMM_XZU}) turns to
\begin{equation}
\begin{split}
&\mathbf{x}_i^{k+1} = {\arg \min}_{\mathbf{x}_i}\left( f_i(\mathbf{x}_i) +{\mathbf{y}_i^{k}}^T(\mathbf{x}_i-\bar{\mathbf{x}}^k_i) + \lambda/2||\mathbf{x}_i-\bar{\mathbf{x}}^k_i||^2_2 \right), \\
&\mathbf{y}_i^{k+1} = \mathbf{y}^k_i + \lambda(\mathbf{x}_i^{k+1} - \bar{\mathbf{x}}^{k+1}_i),
\end{split}
\end{equation}
where $\bar{\mathbf{x}}^k_i = 1/{n_i} \sum_{i=1}^{n_i} \mathbf{x}_i^k$ represents the average of $n$ neighbours of agent $i$. Generally speaking, we gather $\mathbf{x}_i^k$ from outside and scatter $\bar{\mathbf{x}}^k$ to processors, then update $\mathbf{x}_i, \mathbf{y}_i$ in each processor locally in parallel. See details of consensus optimization via ADMM in \cite{Boyd-DistributedOpt}.

\section{Numerical Examples}

In the following we demonstrate the reconstruction performance of the proposed algorithm for quantum state tomography. Two experiments are carried out to show the superior of the proposed algorithm. Consider a quantum state consisting of $q=5$ qubits, its density matrix $\rho$ has size $d \times d$, $d=2^5$. Let the true quantum state as $\rho^*$, $\rho^*$ is generated from normalized Wishart random matrices with form as (\cite{Zyczkowski-GeneratingRan})
\begin{equation}\label{eq:generateRho}
\rho^* = \frac{\mathbf{{\Psi_r}}\mathbf{{\Psi_r}}^*}{\text{tr}(\mathbf{{\Psi_r}}\mathbf{{\Psi_r}}^*)}
\end{equation}
where $\mathbf{\Psi_r}$ is a complex $d \times r$ matrix with i.i.d. complex random Gaussian entries, the denominator is due to the trace $1$ constraint. We construct $\mathbf{A}$ as a $M \times d^2$ sampling matrix whose $M$ rows are chosen randomly without replacement from an $d^2 \times d^2$ matrix whose rows are the set of all vecterized tensor product of Pauli matrices. Maltab R2012b version is used to run the numerical simulations and each value in figures is recorded after averaging 200 experiments.

\begin{figure}[t]
  \centering
  \includegraphics[width=9.2cm]{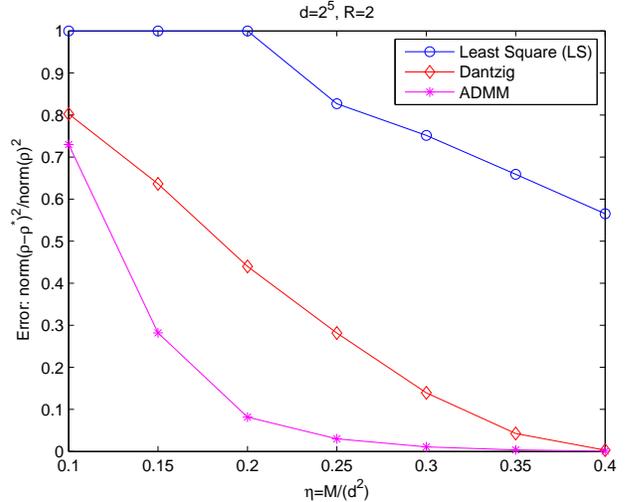}\\
  \caption{The comparison of reconstruction performances of different algorithms, including the least square method in (\ref{eq:rho=argmin}) using cvx toolbox, compressive quantum tomography solving Dantzig in (\ref{eq:Danzig_LASSO}) using cvx toolbox, and compressive quantum tomography solving (\ref{eq:LASSO+ADMM}) using ADMM.}
  \label{ADMM1}
\end{figure}

At first we consider the scenario when the system has small random noises. Here we set $e_i$ in (\ref{eq:y=Arho+e}) satisfies random Gaussian distribution $\mathcal{N}(0, 0.001||\mathbf{\rho}||_2)$. In this case there are two terms in (\ref{eq:rho+I+S}) without $\mathbf{S}$, then the problem is simplified to
 \begin{equation}\label{eq:LASSO+ADMM}
 \begin{split}
 &\text{minimize} ||\mathbf{y}- \mathbf{A} \text{vec}(\rho)||_2 + I_{\mathcal{C}}(\mathbf{z}), \\
 &\text{s.t.} \ \ \rho = \mathbf{z},
 \end{split}
 \end{equation}
where $\mathcal{C}$ represents the low rank Hermitian p.s.d. matrix set. We may update the quadratic term and $I_{\mathcal{C}}(\rho)$ iteratively using ADMM. Specifically, the iteration steps are
\begin{equation}\label{eq:rho_simple}
\begin{split}
\rho^{k+1} &= \text{mat}\left({\arg \min}_{\rho}\left\{ ||\mathbf{y}- \mathbf{A} \rho||_2^2 + \lambda/2 ||\rho- \mathbf{z}^k +\mathbf{u}^k||_2^2\right\}\right), \\
\mathbf{z}^{k+1} &= \Pi_{\mathcal{C}}(\rho^{k+1} + \mathbf{u}^k), \\
\mathbf{u}^{k+1} &= \mathbf{u}^k + (\rho^{k+1} - \mathbf{z}^{k+1}).
\end{split}
\end{equation}
The $\rho$ updating step can be completed by calculating its analytic solution,
\begin{equation}
\rho = \left(\mathbf{A}^*\mathbf{A}+ \lambda \mathbf{I}\right)^{-1} \left( \mathbf{A}^*\mathbf{y} +\lambda(\mathbf{z}^k - \mathbf{u}^k) \right).
\end{equation}
The projection process in step $2$ follows the explanation in Section 3.3 that exploits the shrink or truncated eigenvalue decomposition as in (\ref{eq:rho2}) (\ref{eq:VSigmaV}). In addition, we set rank $r=2$ in the generation of true state $\rho$ in (\ref{eq:generateRho}), $\lambda =1$ in (\ref{eq:rho_simple}). The reconstruction performances are evaluated by the error defined as
\begin{equation}
\text{error} = \frac{||\rho^* - \hat{\rho}||^2_2}{||\rho^*||_2^2},
\end{equation}
where $\rho^*$ and $\hat{\rho}$ denote the true state and the estimate state, respectively. The error is calculated verses the increasing measurement rate $\eta = M/d^2$. Fig.\ref{ADMM1} depicts the reconstruction errors with increasing $\eta$. From Fig.\ref{ADMM1} one can observe that given the low rank information as priori knowledge, the number of measurements is dramatically reduced. Specifically the Dantzig using cvx performs better than the least square approach, and our simplified algorithm using ADMM has smaller errors comparing to Dantzig given the same number of measurements $M$.

\begin{figure}[t]
  \centering
  \includegraphics[width=9.2cm]{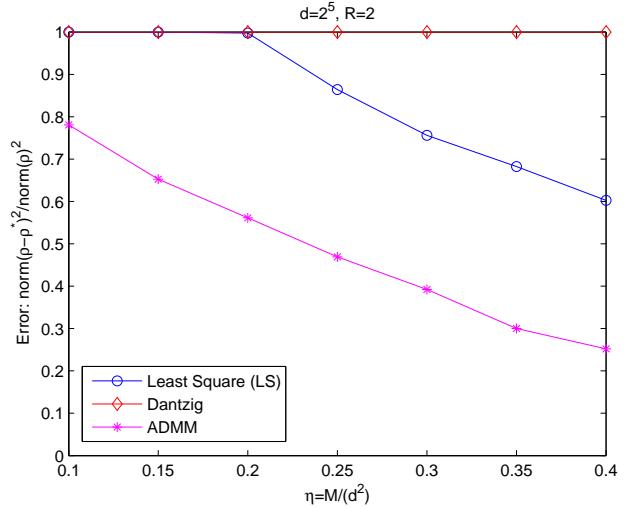}\\
  \caption{The comparison of reconstruction performances with sparse outlier noises, including the least square using cvx toolbox, Dantzig using cvx toolbox, and compressive quantum tomography solving (\ref{eq:LASSO+ADMM}) using ADMM.}\label{ADMM2}
\end{figure}

In the second simulation we add the outlier noises in the density matrix. We set the measurements $\mathbf{y} = \mathcal{A}(\rho+\mathbf{S}) + \mathbf{e}$ where $\mathbf{S} \in d \times d$ has $(0.01d^2)$ nonzero values located randomly with magnitudes satisfying Gaussian distribution $\mathcal{N}(0, 0.1||\rho||_2)$. If the error is larger than $1$, we record it as $1$. The results are demonstrated in Fig. \ref{ADMM2}. Fig. \ref{ADMM2} shows that the Dantzig approach fails under such scenario since the nuclear norm minimization is influenced significantly by the large outliers, however the proposed method may overcome this shortage and still lead to a recovery much better than that of LS method.

\section{Conclusion}

After reviewing several existing algorithms of compressive quantum state tomography and ADMM method, this paper proposes an alternating augmented Lagrangian method for quantum convex optimization problem aiming for recovering pure or nearly pure state with sparse outlier noises. The algorithm updates the density matrix and the sparse estimate noises iteratively and finally obtain a reconstruction result efficiently. Simulations show that the proposed algorithm achieves better recovery accuracy comparing to the conventional least square and compressive Dantzig method with the same number of measurements. For the case of existence of sparse outlier noises, the proposed algorithm beats the Dantzig method due to the fact that the influence of outliers has been reduced.

\section{Appendix}
\begin{prop}
When the quantum state consisting of $q$ qubits is the probabilistic combination of $r$ pure states, then its density matrix $\rho$ with size $d \times d$ has rank not larger than $r$, $d =2^q$.
\end{prop}

\emph{Proof: } The proof is simple however it seldom appears in literatures and lays the foundation of the compressive quantum tomography, so we give a proof here. Suppose the quantum state under consideration
\begin{equation}\label{eq:rho=sum}
 \hat\rho = \sum_{i=1}^{r} p_i |\psi_i \rangle \langle \psi_i|,
\end{equation}
which means the quantum system may be found in state $| \psi_i \rangle$ with probability $p_i$, $i=1,2, \cdots,r$. If we concatenate the column vectors $| \psi_i \rangle$ as a matrix, then (\ref{eq:rho=sum}) is equivalent to
\begin{equation}\label{eq:rho=matrix}
\begin{split}
& \hat\rho = \mathbf{\Psi_r} \cdot \mathbf{\Psi_r}^*, \text{where} \\
 &  \mathbf{\Psi_r}= \left[ \sqrt{p_1} |\psi_1 \rangle,  \sqrt{p_2} |\psi_2 \rangle , \cdots, \sqrt{p_r} |\psi_r \rangle \right],
\end{split}
\end{equation}
$\mathbf{\Psi_r}$ is of size $d \times r$. Thus the density matrix $\hat\rho$ has rank at most $r$ due to the rank property of multiplication of two matrices.

\begin{deft}[$\mathbf{ Rank \ RIP}$]\cite{RECHT-Guaranteed,Liu-Universal-Pauli}
The $\mathcal{A}$ satisfies the rank restricted isometry property (RIP) if for all $d \times d$ $\mathbf{X}$, we have
\begin{equation}
(1-\delta)||\mathbf{X}||_F \leq ||\mathcal{A} (\mathbf{X})||_2 \leq (1+\delta)||\mathbf{X}||_F
\end{equation}
where some constant $0 < \delta <1$.
\end{deft}

\begin{prop}
When we formulate the measurement process as equations in (\ref{eq:y=Arho+e}), and the observable $\mathbf{O}_i$ are the tensor/Kronecker product of a series of complex and unitary elemental $2 \times 2$ Pauli matrices $\mathbf{P}_i$ chosen from the four possibilities randomly,
\begin{equation}\label{eq:Pauli}
\begin{split}
\mathbf{I}_2&=\left( \begin{array}{cc} 1  & 0 \\ 0 & 1   \end{array}\right),  \mathbf{\sigma}_x =\left( \begin{array}{cc} 0  & 1 \\ 1 & 0   \end{array}\right), \\
\mathbf{\sigma}_y&=\left( \begin{array}{cc} 0  & -i \\ i & 0   \end{array}\right),  \mathbf{\sigma}_z =\left( \begin{array}{cc} 1  & 0 \\ 0 & -1   \end{array}\right).
\end{split}
\end{equation}
Then sampling operator $\mathcal{A}$ satisfies the rank RIP and we are able to recover the rank-$r$ density matrix $\rho$ by using number of measurements $m \leq c\cdot rd \log^6 d$ for some absolute constant $c$ with high probability.
\end{prop}
\emph{Proof:} The details of the proof can be found in \cite{Gross-RecoveringLow-Rank,Liu-Universal-Pauli}.
\bibliographystyle{plain}
%\bibliography{KezhiRefNew46}

%\bibliography{IEEEabrv}

%\section*{Acknowledgment}


\end{document}